# Carrier doping to pseudo-low-dimensional compound $La_2RuO_5$

Masatomo Uehara*, Kenich Ashikawa, Yoshimasa Aka and Yoshihide Kimishima

*Department of Physics, Faculty of Engineering, Yokohama National University, Hodogaya-ku Yokohama 240-8501, Japan*

**Abstract**

Hole carrier doping has been tried to pseudo-low-dimensional material $La_2RuO_5$ by substituting $La^{3+}$ with $Cd^{2+}$. Single phased samples of $La_{2-x}Cd_xRuO_5$ with $x$ up to 0.5 have been successfully obtained and also high pressure $O_2$ annealing has been performed to the $x$=0.5 sample. Although the formal ionic state of Ru is expected to increase from 4+ (at $x$=0) to 4.5+ (at $x$=0.5), the magnetic and electrical properties show no significant changes in as-sintered samples. In contrast, high pressure $O_2$ annealed $x$=0.5 samples show a little reduction of electrical resistivity and the decrease of thermoelectric power at 260 K. From these results, it can be speculated that the doped carriers are mostly compensated by oxygen deficiency in as-sintered samples.

---

* Corresponding author

Address : Department of Physics, Faculty of Engineering, Yokohama National University, 79-1 Tokiwadai, Hodogaya-ku Yokohama 240-8501, Japan

E-mail : uehara@ynu.ac.jp

Tel & Fax : +81-45-339-4187

## Introduction

The magnetic and electrical properties of $La_2RuO_5$ were first reported by P. Khalifah et.al. [1]. The crystal structure at above 165 K is a two-dimensional structure with corrugated Ru-O layers spreading over *a*-*c* plane separated by La-O slabs along *b*-axis as shown in figure 1 [1-3]. Below 165 K structural phase transition takes place, then the nearest neighbor Ru-Ru bond length, which is almost identical to be 3.98 Å above 165 K both in *a*-*b* plane and along *c*-axis, changes into two long (4.05 Å) and short (3.87 Å) ones in *a*-*b* plane shown in fig. 1 [1-3]. Therefore, low temperature phase can be regarded as two-leg ladder structure with legs running along *c*-axis. In addition, the bond length alternation also occurs along *c*-axis below 165 K (3.92 Å and 4.04 Å), so that it is better to view the structure as two-leg ladder consisting of alternative one-dimensional chains.

Accompanied by this structural phase transition, the semiconducting activation energy of the polycrystalline sample slightly increases from 0.23 eV to 0.32 eV, reflecting the reduction of dimensionality [1]. In contrast to this, the magnetic properties dramatically change. At 165 K magnetic to non-magnetic transition occurs. The origin of this magnetic transition has been interpreted by two different scenarios [1,4,5]. One is the phase transition of $S$=1 intermediate spin state of $Ru^{4+}$ to $S$=0 low spin state caused by local distortion of $RuO_6$ octahedron and by subsequent orbital ordering [1]. In the other scenario, even though there exist the local distortion of $RuO_6$ octahedron, Ru ions retain $S$=1 spins and undergo the spin singlet transition caused by spin ladder formation below 165 K [4,5]. If the latter scenario is the case, superconductivity might be expected at metallic state induced by carrier doping. In this paper, we report electrical and magnetic properties of hole-carrier doped $La_{2-x}Cd_xRuO_5$.

**Experimental**

A series of samples with nominal composition $La_{2-x}Cd_xRuO_5$ was synthesized from $La_2O_3$, Cd, $RuO_2$ powders. The powder mixtures were pressed into pellets and calcined at 900 °C for 5 h in air. After calcinations, the samples were sintered at 1050 °C for 10 h in air two times with intermediate grindings. It is worth mentioning that $La_{2-x}Cd_xRuO_5$ can not be synthesized in $O_2$ gas but synthesized in air. For $x$=0.5 sample, high pressure $O_2$ (HP-$O_2$) annealing was applied by using a cubic anvil high-pressure apparatus. The $x$=0.5 sample was sealed in Au cell together with $KClO_4$ that works as oxygen supplier. Annealing treatments were done by three different methods. 1) 0.3g $La_{1.5}Cd_{0.5}RuO_5$+ 0.006g $KClO_4$ was annealed in 530 °C for 2 h at 1 GPa. (sample name : 0.5A). 2) 0.3 g $La_{1.5}Cd_{0.5}RuO_5$+ 0.011g $KClO_4$ was annealed in 530°C for 4h at 1GPa. (sample name : 0.5B). 3) 0.3 g $La_{1.5}Cd_{0.5}RuO_5$+ 0.011g $KClO_4$ was annealed in 530 °C for 12 h at 1 GPa. (sample name : 0.5C). Annealing condition become stronger from 0.5A to 0.5C. X-ray diffraction patterns were measured by CuKα radiation. Resistivity measurement was done by standard four-probe method with the $^4$He cryostat. Magnetization measurement was performed with zero field cooling (ZFC) and field cooling (FC) by using Quantum Design SQUID magnetometer. Thermoelectric power was measured by the standard steady-state method.

## Results and discussion

The powder x-ray diffraction patterns for $La_{2-x}Cd_xRuO_5$ taken at room temperature are shown in figure 2. Almost all peaks are indexed by assuming monoclinic crystal structure. $LaRuO_3$ was detected as an impurity phase. KCl peaks seen in the x-ray pattern of 0.5B and 0.5C samples come from the $KClO_4$ used by oxygen supplier. Lattice parameters are determined with JADE program and figure 3 shows the calculated lattice parameters of $La_{2-x}Cd_xRuO_5$ as a function of Cd content $x$. The $a$- and $b$-axis increase and $c$- axis slightly decrease with increasing $x$. Based on these changes of lattice parameters at high temperature phase, it might be supposed that the same changes of lattice parameters as a function of $x$ can be adopted for the low temperature phase. Therefore, as $x$ increase, both inter and intra ladder distances are shorten corresponding to the shrinkage of $a$ and $b$-axis, and the distance parallel to the ladder is enlarged reflecting the increase of $c$-axis. $\beta$ shows no significant change against the Cd doping and HP-$O_2$ annealing. For HP-$O_2$ annealed samples, $a$- and $b$- axis seem to slightly increase, comparing with as-sintered sample of $x$=0.5. However, the amount of changes is almost within error bar so that it is not clear if lattice constants change or not by HP-$O_2$ annealing.

The temperature dependence of magnetic susceptibility ($\chi$) of $La_{2-x}Cd_xRuO_5$ is shown in figure. 4. At around 165 K, sudden drops of susceptibility are seen, indicating the paramagnetic to non-magnetic transition. The transition temperatures show no significant change in all measured samples with different doping level. Below 50 K, the upturn of $\chi$ is seen and $\chi$ shows the deviation between ZFC and FC (shown only $x$=0.2 sample) data, suggesting the existence of not only Curie-term but also ferromagnetic or spin glass-like component. We have supposed that that this behavior is not intrinsic one of $La_{2-x}Cd_xRuO_5$ but comes from undetectable amount of $La_{3.5}Ru_4O_{11}$ by x-ray. Inset of Fig. 4 shows temperature dependence of magnetic susceptibility $\chi$ of $La_{3.5}Ru_4O_{11}$. Below 50 K the $\chi$ of $La_{3.5}Ru_4O_{11}$ behaves similar to that of $La_{2-x}Cd_xRuO_5$. Similar behavior in $\chi$ below 50 K has been reported [6]. For 0.5A, 0.5B and 0.5C samples, Curie terms at low temperature seem to be enhanced comparing to as-sintered samples. This might be attributed to the free spins coming from the singlet broken by hole carriers produced from additional oxygen introduced by HP-$O_2$ annealing. The $x$=0 sample also shows the enhanced Curie terms at low temperature and in the case this behavior is probably due to bad sample quality (see the x-ray data of fig. 2).

Effective magnetic moments ($p_{eff}$) are calculated from magnetization curves above 180 K by the fitting procedure assuming Curie-Weiss law and summarized as a function of $x$ in figure.5. For as-sintered samples, $p_{eff}$ does not show systematic changes with $x$ and their values are within approximately 2.7~2.75, consistent with the value of $S$=1 $Ru^{4+}$ ion. The solid line in fig. 5 represents the calculated values by assuming that the doped Ru sites behave as $S$=3/2 localized spins ($Ru^{5+}$ $4d^3$). Comparing experimental values with calculated ones, $p_{eff}$ seems to be constant as if the carriers are not doped in spite of $Cd^{2+}$ substitution. In HP-$O_2$ annealed samples, $p_{eff}$ seems to slightly increase comparing with as-sintered $x$=0.5 sample. This might suggest that the HP-$O_2$ annealing produces some of $Ru^{5+}$ ions.

Figure 6 shows the temperature dependence of electrical resistivity ($\rho$) of $La_{2-x}Cd_xRuO_5$. All samples show semiconducting behavior and some of the samples shows the kinks around 160 K corresponding to the structural phase transition. Inset of fig. 6 shows the electrical resistibvity values at 273 K ($\rho_{273\ K}$) as a function of $x$. For as-sinterd sample, $\rho_{273\ K}$ fluctuates between ~$10^2$ and ~$10^3$ Ohm·cm without systematic change depending on $x$, so that it can be supposed that the $\rho_{273\ K}$ is almost constant regardless of $x$ and the fluctuation probably comes from the variation of sample quality. However, HP-$O_2$ annealed samples, the resistivity decrease from that of as-sintered $x$=0.5 sample, indicating that HP-$O_2$ annealing effectively works for carrier doping.

Figure 7 shows the temperature dependence of thermoelectric power ($S$) for $La_{2-x}Cd_xRuO_5$. with $x$=0.1, 0.3, 0.5 and 0.5C samples. In thermoelectric power measurement, quantitatively reliable data can be obtained even in polycrystalline sample because the measurement is performed under the situation where the thermal and electric current do not flow through the sample so that the measurements is not so affected by grain boundary. The sign of thermoelectric power is plus, indicating the carrier is hole. The value of $S$ slightly decreases with increasing $x$. However, by HP-$O_2$ annealing, $S$ largely decreases as seen in 0.5C sample (see inset of fig.7). This implies that the carrier number is not so much increased by Cd doping in as-sintered samples but effectively increased by HP-$O_2$ annealing, suggesting that there is the oxygen deficiency in as-sintered samples and HP-$O_2$ annealing fills the oxygen vacancies.

From the magnetization and resistivity data, it has been observed that the magnetic and electrical properties does not change so much by Cd doping for as-sintered samples even though carriers are though to be doped by Cd substitution. Regarding this result, it is possible to assume that there is the oxygen deficiency that compensates the hole-carrier introduction by Cd-doping. Experimentally, by HP-$O_2$ annealing, the resistivity is slightly decreased and the enhanced Curie term is observed at low temperature probably caused by free spins from the singlet broken by doped holes. Thermoelectric power measurements also show that the HP-$O_2$ annealing effectively increase the carrier number. These experimental results can be explained by filling the oxygen vacancies and consequent increase of hole carrier number with HP-$O_2$ annealing. Moreover, as mention in experimental section, $La_{2-x}Cd_xRuO_5$ can not be synthesized in $O_2$ gas but in air. This might suggest that $La_{2-x}Cd_xRuO_5$ is stabilized as the equilibrium phase with some amount of oxygen deficiency. Actually, the thermogravimetric experiment of $La_2RuO_5$ up to 800 ℃ in air has observed a weight loss due to oxygen deficiency. Furthermore, in the thermogravimetric analysis of $La_2RuO_5$ in Ar90%+$H_2$10% mixed gas, it has been shown that the oxygen defect phase $La_2RuO_{4.6}$ is present as a stable phase [7]. From the facts mentioned above, it can be expected that $La_{2-x}Cd_xRuO_5$ is also robust against oxygen deficiency, and that the material tends to easily have oxygen deficiency. At present annealing condition, the metallic conduction has not been achieved. As another obstructive factors for carrier doping 1) it is possible to assume that the Cd-4*d* orbital hybridizes with Ru-4*d* orbital so that the hole carriers are partly compensated and hole doping is not so smooth with Cd substitution. 2) There is the possibility that the Cd volatilizes during the sintering process, therefore Cd content is smaller than the nominal composition. It is necessary to perform quantitative analysis.

## Conclusion

$La_{2-x}Cd_xRuO_5$ has been successfully synthesized up to $x$=0.5. Magnetization and electrical measurements show no significant and systematic change even though hole carriers are thought to be doped by Cd-substitution. From these results, the existence of oxygen deficiency is suggested. Magnetization, electrical resistivity and thermoelectric power measurements have shown that the HP-$O_2$ annealing effectively increase the carrier concentration. These experimental results can be explained by filling the oxygen vacancies and consequent increase of hole carrier number with HP-$O_2$ annealing. However, metallic conduction and probable superconductivity have not been achieved at present stage. More effective annealing condition for carrier doping is now being explored.

This work was partly supported by a Grant-in-Aid for Scientific Research from The Ministry of Education, Culture, Sports, Science and Technology, Japan and by Research Institute of Yokohama National University.

**Figure captions**

Figure 1 : Schematic view of *a*-*b* plane cross section for $La_{2-x}Cd_xRuO_5$. Left and right side views show high and low temperature structures, respectively.

Figure 2 : Powder x-ray diffraction patterns for $La_{2-x}Cd_xRuO_5$

Figure 3 : Lattice parameters of $La_{2-x}Cd_xRuO_5$ as a function of Cd content $x$.

Figure 4 : Temperature dependence of $\chi$ for $La_{2-x}Cd_xRuO_5$. For $x$=0.2 sample, the FC data is shown in addition to ZFC data. Inset shows the temperature dependence of $\chi$ for $La_{3.5}Ru_4O_{13}$.

figure.5 : Cd content $x$ dependence of effective magnetic moments $p_{\mathrm{eff}}$ calculated from magnetization curves above 180 K by the fitting procedure assuming Curie-Weiss law. The solid line represents the calculated values by assuming that the doped Ru sites behave as $S$=3/2 localized spins ($Ru^{5+}$ 4$d^3$).

Figure 6 : Temperature dependence of $\rho$ down to 10 K for $La_{2-x}Cd_xRuO_5$. Inset shows the $\rho$'s of various $x$ at 273 K.

Figure 7 : Temperature dependence of the thermoelectric power $S$ for $La_{2-x}Cd_xRuO_5$. Inset shows the $S$ values with $x$=0.1, 0.3, 0.5 and 0.5C samples at 260 K.

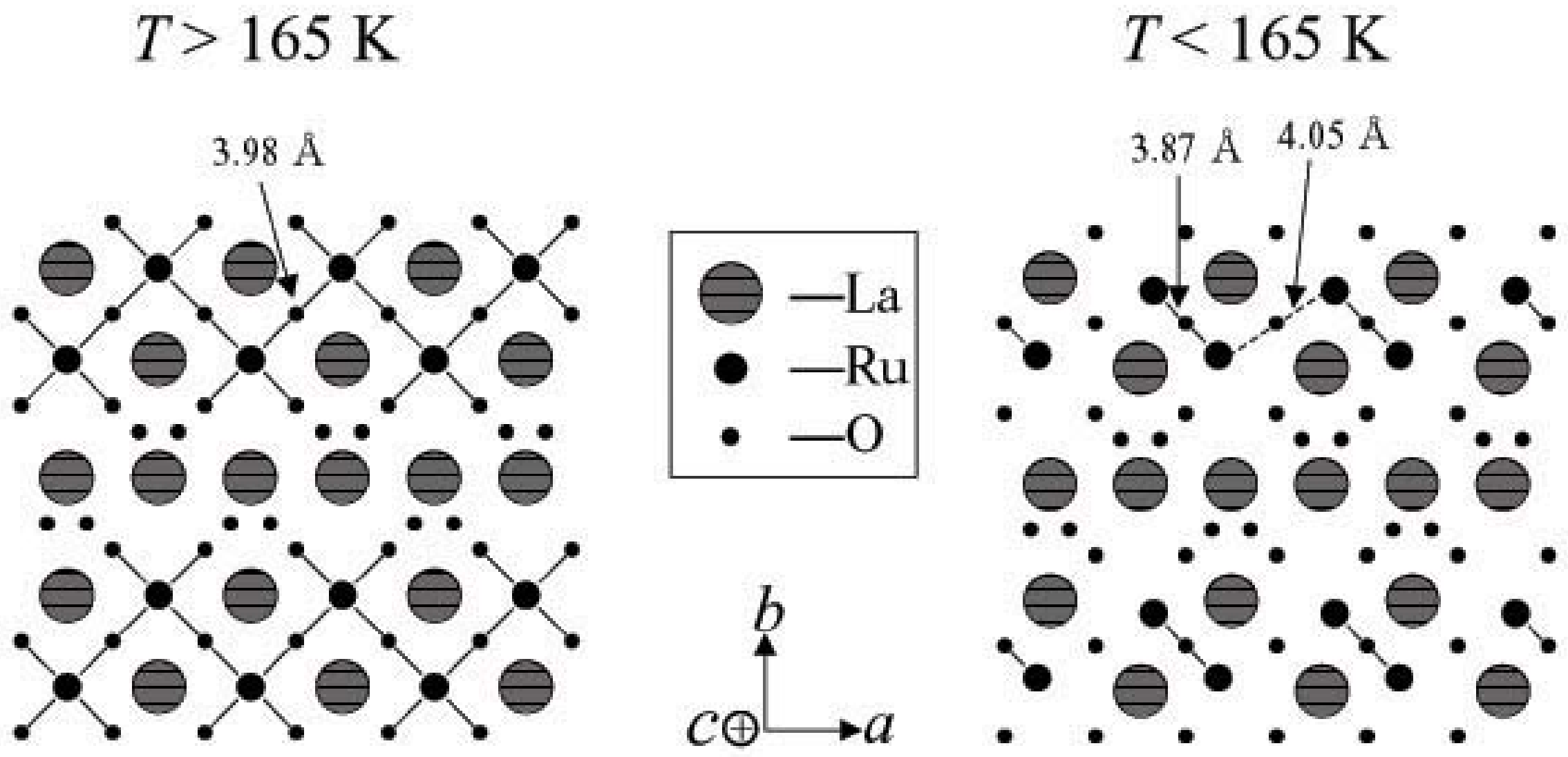

Fig. 1

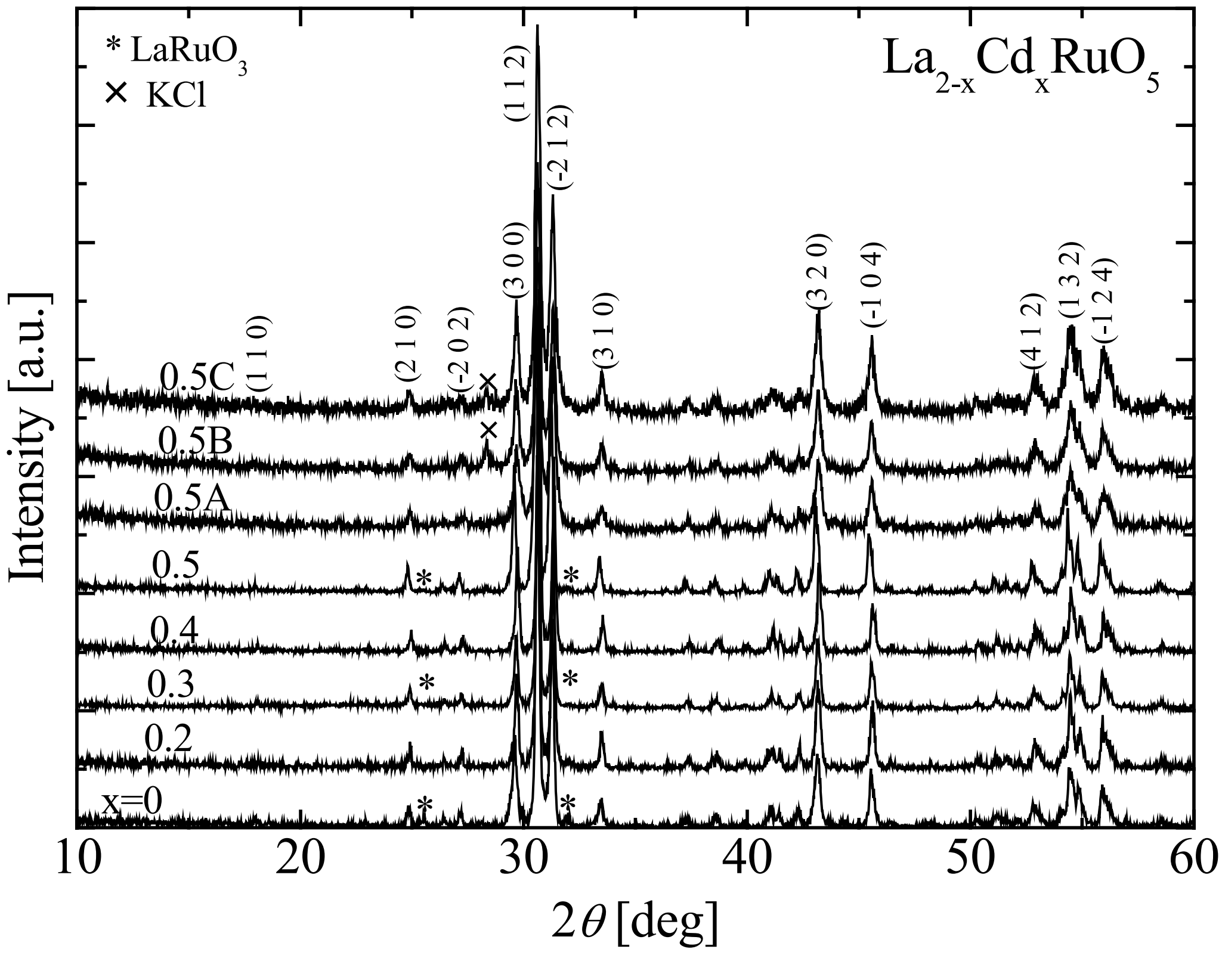

Fig. 2

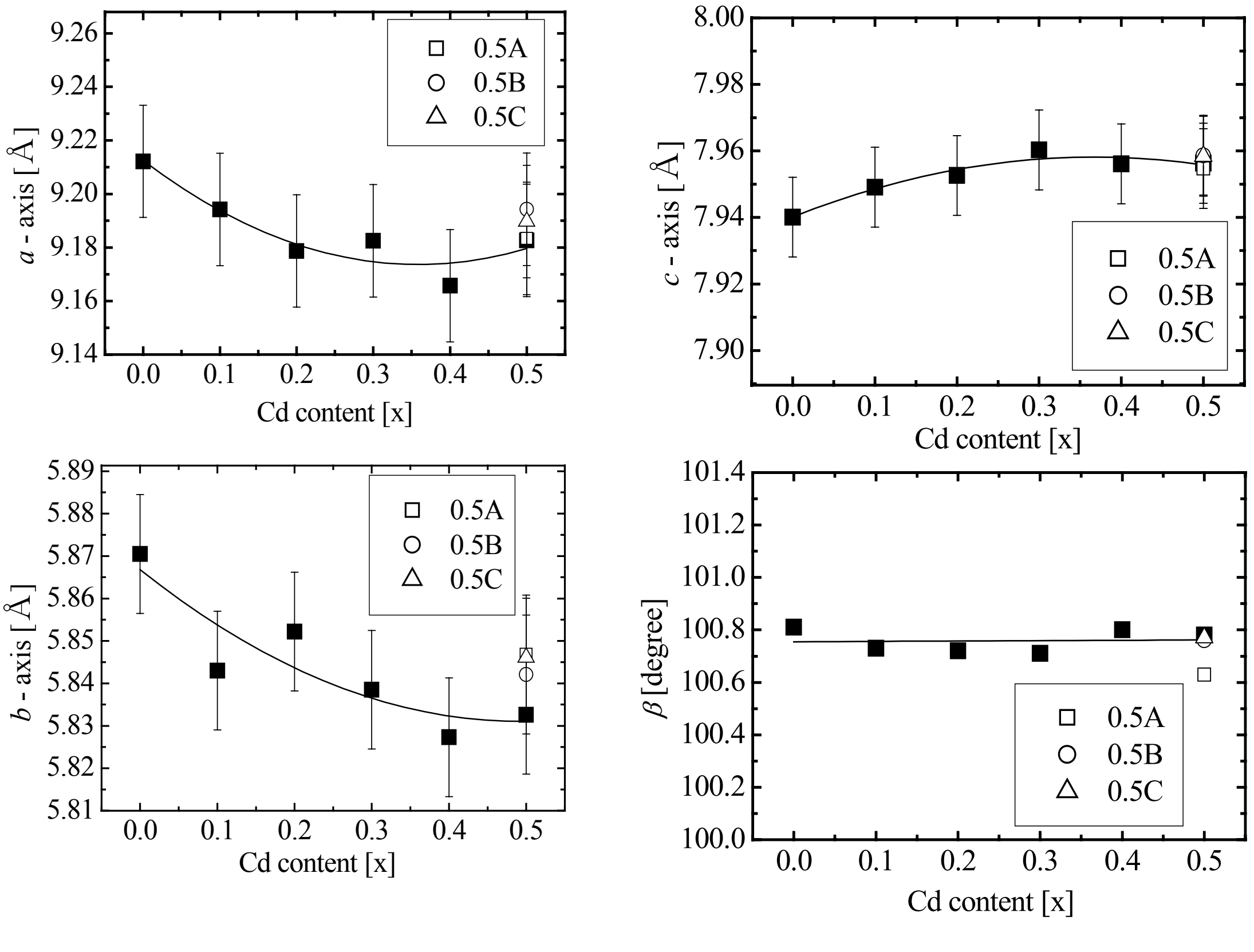

Fig. 3

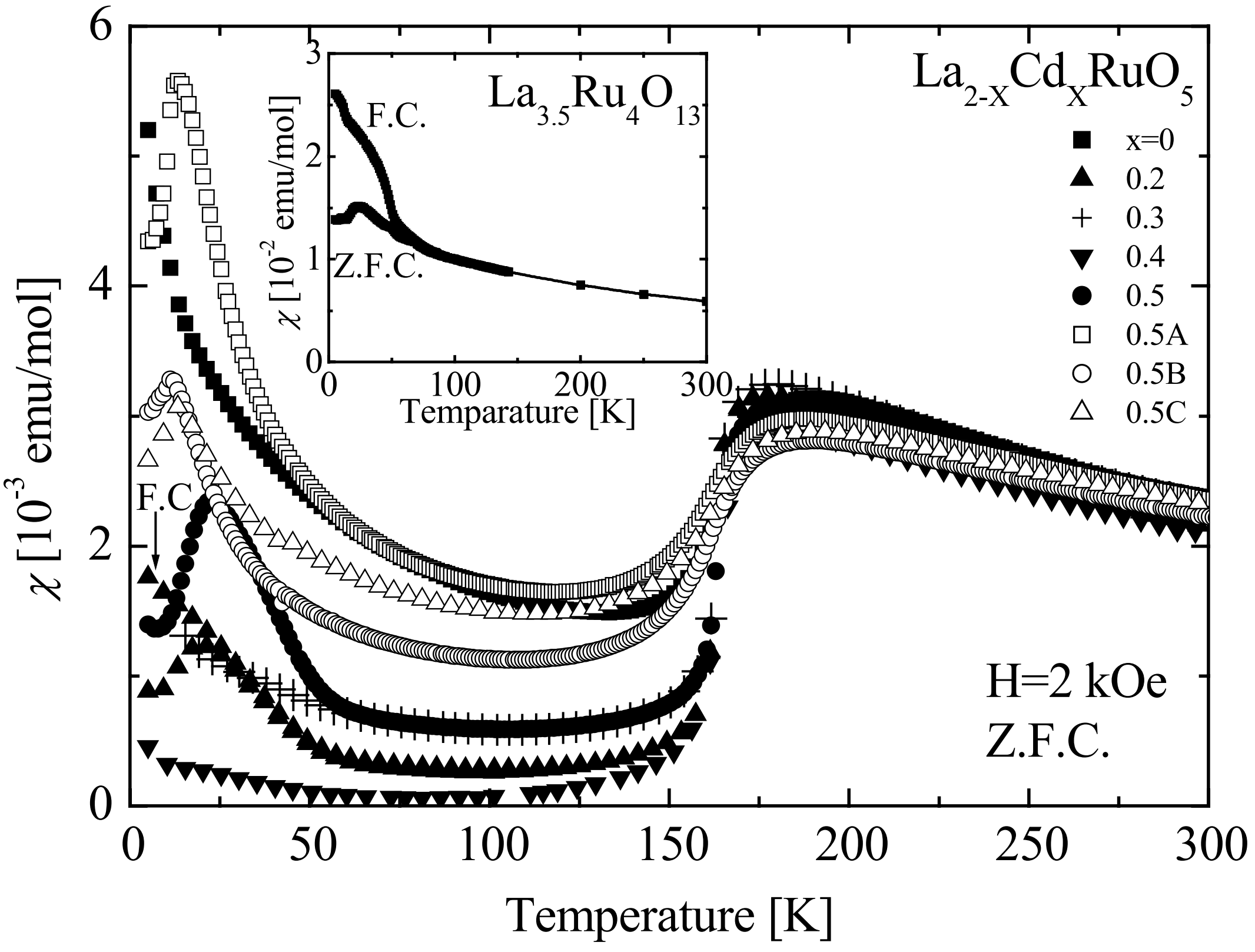

Fig. 4

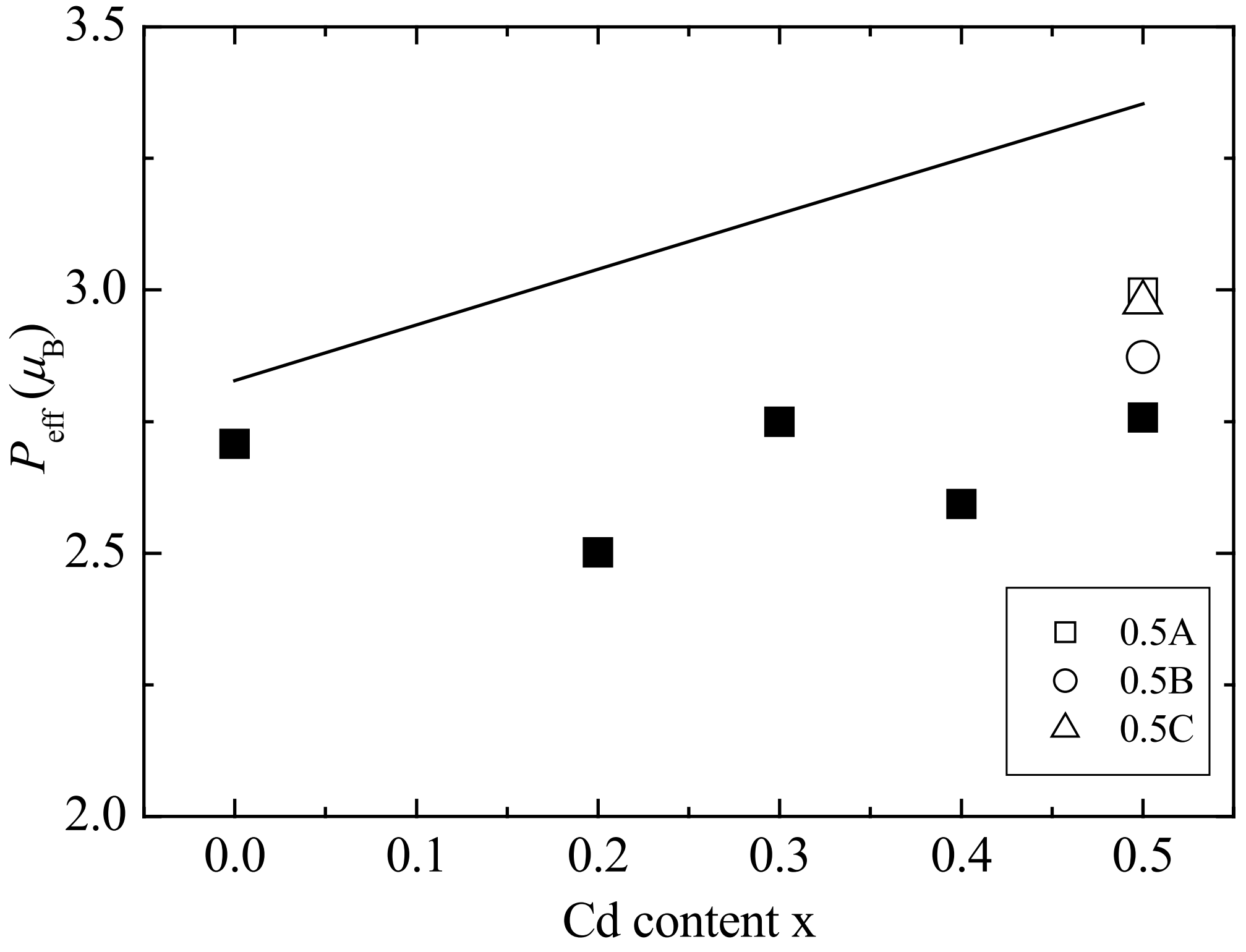

Fig. 5

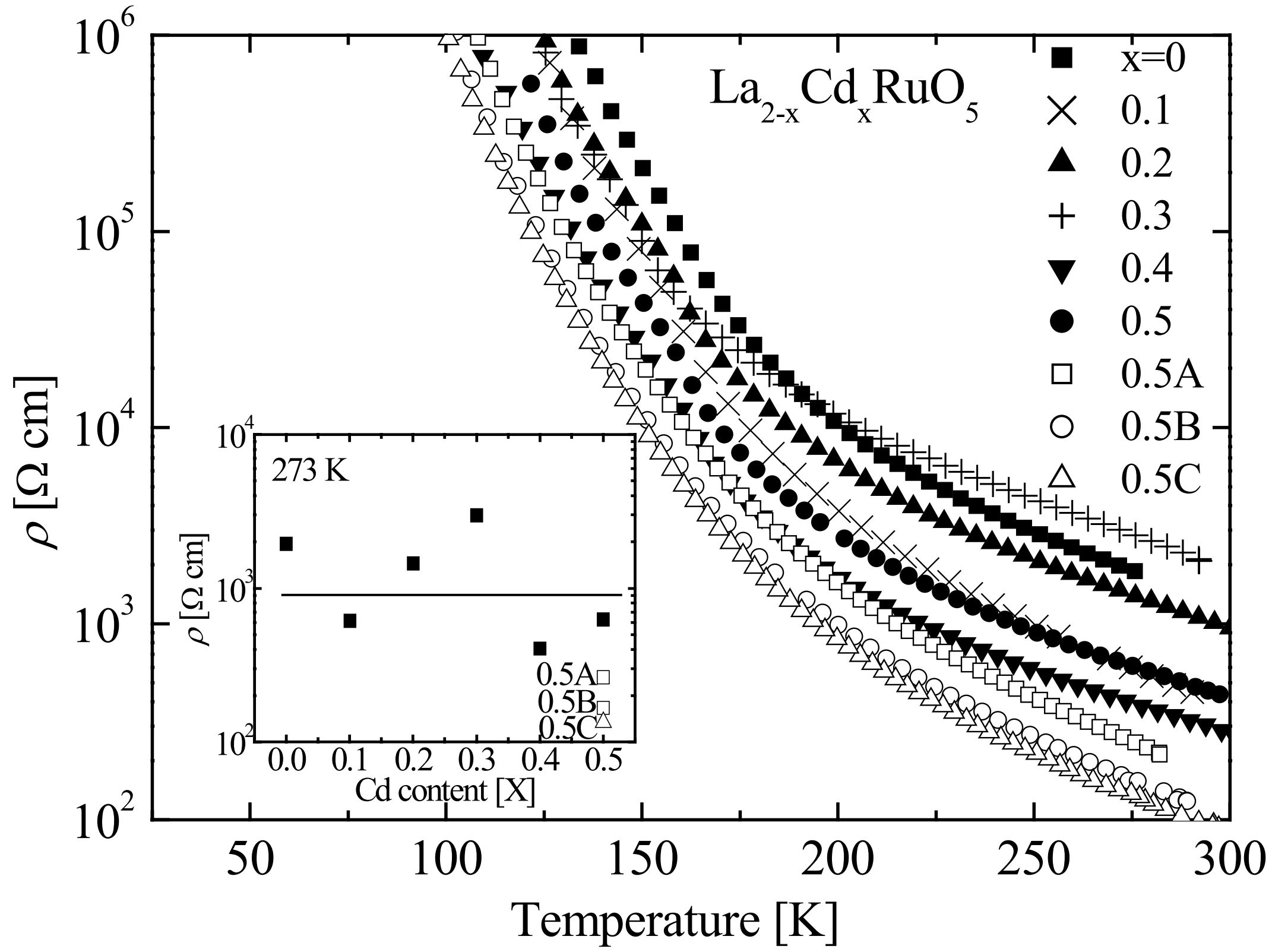

Fig. 6

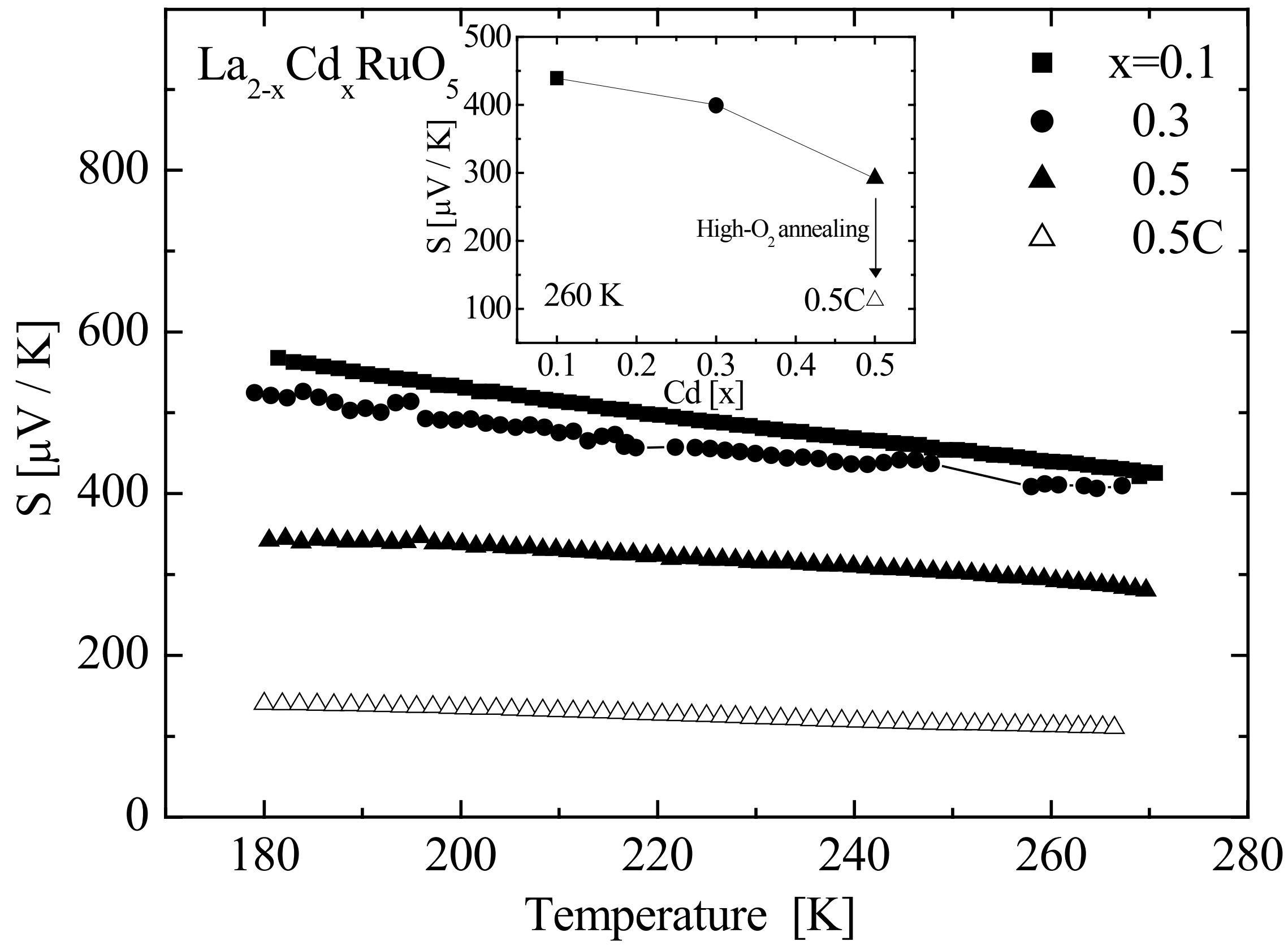

Fig. 7